\documentclass[aps,pra,10pt,twocolumn,showpacs,superscriptaddress]{revtex4-2}

\usepackage{physics}
\usepackage{graphicx}
\usepackage[colorlinks=true,linkcolor=blue,citecolor=red,plainpages=false,pdfpagelabels]{hyperref}
\usepackage{times}

\def\EQ#1{\begin{eqnarray}#1\end{eqnarray}}
\newcommand{\map}[1]{\mathcal{#1}}

\begin{document}

\title{\Large{\textbf{Information Bleaching, No-Hiding Theorem and Indefinite Causal Order}}}


\author{Abhay Srivastav}\email{abhaysrivastav@hri.res.in}
\affiliation{Harish-Chandra Research Institute, Chhatnag Road, Jhunsi, Prayagraj 211019, India}
\affiliation{Homi Bhabha National Institute, Training School Complex, Anushaktinagar, Mumbai 400094, India}
\author{Arun Kumar Pati}\email{akpati@hri.res.in}
\affiliation{Harish-Chandra Research Institute, Chhatnag Road, Jhunsi, Prayagraj 211019, India}
\affiliation{Homi Bhabha National Institute, Training School Complex, Anushaktinagar, Mumbai 400094, India}

\begin{abstract}
The information bleaching refers to any physical process that removes quantum information from the initial state of the physical 
system. The no-hiding theorem proves that if information is lost from the initial system, then it cannot remain in the bipartite quantum correlation and must be found in the remainder of the Hilbert space. We show that when hiding map acts on the input state in the presence of indefinite causal order, then it is possible to hide quantum information in the correlation. One may ask, does it then violate the no-hiding theorem? We analyse this question and argue that in the extended Hilbert space, it will still respect the no-hiding theorem. We also discuss how to mask quantum information using superposition of two hiding maps. Our results can have interesting implications in preserving the fidelity of information, preservation of quantum coherence and work extraction in the presence of two hiding maps with indefinite causal order. Furthermore, we apply the hiding maps in the presence of indefinite causal order on half of an entangled pair and show that entanglement cannot be preserved. 
 Finally, we discuss that even though quantum entanglement is destroyed, the entanglement fidelity under indefinite causal order is non-zero and can approach close to one.
\end{abstract}

\maketitle

\section{Introduction}

The linearity and the unitarity are the two fundamental aspects of quantum mechanics.
These two tenets have profound consequences on the 
structure of quantum theory and quantum information. The no-cloning theorem \cite{ww} and the no-deleting theorem \cite{akp} 
essentially follow from the linearity and the unitarity. Interestingly, these two theorems also show that quantum information can neither be created nor be destroyed
in any physical process \cite{rj,horo}. 
When a quantum system interacts with an external world, it may loose its information leading to loss of coherence and purity. If the 
original information is completely lost from the final state, we call the physical process as the bleaching process. Now, one may wonder 
if the quantum information is bleached out, then where has the information gone? The no-hiding 
theorem \cite{sam}  
answers this question. The precise statement of the no-hiding theorem is that if the original information is lost, then it cannot remain 
in the bipartite correlation and must be moved to some subspace of the environment. Thus, any physical process that tends to destroy information, has to respect the no-hiding theorem. This has applications in state randomization, thermalization and even black hole evaporation. The original motivation for the no-hiding theorem actually stemmed from the black hole information loss and it was found that quantum information cannot remain in the correlation between the inside and the outside state of black hole. The theorem was recently tested using the technique of nuclear magnetic resonance (NMR) where state randomization was taken as an example of the bleaching process and the lost information was fully recovered from the ancillary state \cite{anil}. This experiment in a sense, for the first time, demonstrated the notion of conservation of quantum information. 

Quantum mechanics is usually formulated in a background of fixed causal order between events. However, recent progresses have shown that the structure of causality is much more rich, thus allowing for events to be connected by indefinite causal order \cite{i2}. An example of an inherent indefinite causal order phenomena is quantum switch \cite{i1,i3,i4,i5,comp} which is nothing but the superposition of the causal order of quantum operations aided by a control qubit.
It has been recently demonstrated that a superposition of the order of quantum channels via quantum switch offers a non trivial resource in information theory. This can give rise to non trivial advantages on the top of quantum advantages (with fixed causal order) in quantum computing \cite{comp}, reducing the communication complexity \cite{compl1,compl2}, quantum communication \cite{comm, mitra}, quantum teleportation \cite{chira1}, quantum thermometry \cite{chira2}, remote coherence creation \cite{jas} and in the task of channel discrimination \cite{chd1,chd2}. In recent years, the notion of indefinite causal order has also been experimentally realized \cite{exp1,exp2} and has been generalised to superposition of $N$ channels \cite{sazim}.

In this paper, we consider the information bleaching process and apply the no-hiding theorem in the presence of indefinite causal order. In the absence of indefinite causal order, the hiding map will bleach out complete information. However, 
 when the hiding map acts on the input state in the presence of indefinite causal order, then it is possible to hide quantum information in the correlation. One may ask, does it then violate the no-hiding theorem? We analyse this question and argue that in the extended Hilbert space, it will still respect the no-hiding theorem. 
 Our result has interesting connection to the no-masking theorem \cite{modi,liu,li,qiao,bao,liang} which states that an arbitrary quantum state cannot be masked by a unitary process, i.e., we cannot take a pure state and map it to pure entangled state in a larger Hilbert space with the condition that individual subsystems contain no information about the input. However, we will show that using the indefinite causal order it is possible to superpose two channels such that any arbitrary state can be mapped to a bipartite state and the individual subsystems do not contain any information about the input. Though, this encoding is not unitary, but a general completely positive trace preserving map (CPTP).
 Furthermore, we show how much information can remain in the correlation when a conditional measurement is performed on the control qubit. To answer this question, we find the fidelity of information  as well as
quantum coherence in the conditional state. We find that it is possible to achieve fidelity of the final state more than the classical value and maintain non-zero coherence in the presence of indefinite causal order.
This may have interesting implications for the black hole information loss. If two black holes are in the superposition of indefinite causal order, then it may be possible that information is neither inside nor outside the black hole and can reside completely in the 
correlation. Next we show the thermodynamic advantage of indefinite causal order by showing that it can help in work extraction even when a single hiding map gives zero extractable work. We also apply the hiding map in the presence of indefinite causal order to half of an entangled pair and find that even though 
quantum information of an arbitrary pure state can be preserved with fidelity more than the classical value, quantum entanglement is not preserved.
The final output state is a completely separable state. 
This suggests that positive coherent information cannot be sent across such a channel even if indefinite causal order is used.
 We show that even though quantum entanglement is destroyed, the entanglement fidelity with two uses of the hiding map under indefinite causal order is non-zero and can approach close to one for some choice of the control qubit.
Thus, our results throw new lights on the nature of preservation of quantum coherence and entanglement in the presence of indefinite causal order and bleaching process.

The paper is organised as follows. In section II, we provide the basic notions about the bleaching process, the hiding map along with the no-hiding theorem and indefinite causal order. In section III, we present our main findings, i.e., we show how to keep quantum information in the correlation of bipartite state using the hiding map in the presence of indefinite causal order. In section IV, we discuss the bleaching process when the hiding map acts on half of an entangled pair and show that it is not possible to preserve entanglement in the presence of indefinite causal order. Finally, we conclude in section V.

\section{Preliminaries}
In this section, we discuss the preliminary notions about the bleaching process, the no-hiding theorem and the indefinite causal order.

\subsection{Information Bleaching Process}
The bleaching process refers to any physical process that completely removes information from the final state of the system and the corresponding mathematical map is called the hiding map. It is a completely positive and trace preserving (CPTP) map  that transforms an arbitrary input state 
$\rho = \ketbra{\psi}{\psi}$ to a fixed output state $\sigma$, i.e., we have 
\begin{equation}
\mathcal{E}: \rho \rightarrow \sigma,\label{equation:1}
\end{equation}
where $\sigma$ has no dependence on $\rho$.
Let the spectral decomposition of $\sigma$ be $\sum_{k}p_{k}\ketbra{k}{k}$, where $p_{k}$ are the eigenvalues and $\{\ket{k}\}$ are the eigenvectors of $\sigma$. The simplest example of bleaching process is state randomization. If $p_{k}= 1/d$ $\forall k$, where $d$ is the dimension of $\sigma$, hiding process turns out to be state randomization. Similarly, if $p_{k} = 1$ for $k = d$ (where the dimension of the input Hilbert space is $d-1$) and 0 otherwise, then $\mathcal{E}: \rho \rightarrow \ketbra{d}{d} $ and it becomes the erasure process \cite{wilde}. If we consider the output state $\sigma$ as a thermal state, i.e., $\sigma= e^{-\beta H}/ \Tr(e^{-\beta H}) = \sum_k p_k \ketbra{k}{k}$, where
$p_k = e^{-\beta E_k}/Z$,  $Z= \sum_k e^{-\beta E_k}$ and $\{\ket{k}\}$ are the energy eigenbasis of the bath Hamiltonian, then the bleaching process can represent thermalization. Thus, the bleaching process generalises  and unifies several physical processes that bleaches the information of an input system.\\
Now if we ask, where does this lost information go? The no-hiding theorem provides the answer: information about the input system if bleached out from its output, moves entirely to the remainder of the ancilla. No information about the system remains hidden in the correlations between its output and the ancilla. This is a remarkable consequence of the linearity and the unitarity property of quantum mechanics.\\\\
\noindent
{\bf The No-Hiding Theorem:} If $\ketbra{\psi}{\psi} \rightarrow \sigma$ under a CPTP map with $\sigma$ being a fixed state independent of the input, then in the enlarged Hilbert space ${\cal H}_S \otimes {\cal H}_A$, the unitary process must map $\ket{\psi}_S \rightarrow \ket{\Psi}_{SA} $ such that $\ket{\Psi}_{SA} = \sum_k \sqrt p_k \ket{k}_S \otimes (\ket{q_k} \otimes \ket{\psi} )_A$  with 
$\Tr_A (\ketbra{\Psi}{\Psi}) = \sigma$.\\\\
Using the Choi-Kraus representation theorem \cite{wilde,choi} any CPTP map $\mathcal{E}$ from the set of density operators of the input Hilbert space to the set of density operators of the output Hilbert space, can be written in the form $\mathcal{E}(\rho)= 
\sum_{\mu } F_{\mu}\rho F_{\mu}^\dagger$ where $\{ F_{\mu} \}$ are the Kraus operators of $\mathcal{E}$ such that $\sum_{\mu}F_{\mu}^\dagger F_{\mu}=I$ and $\rho$ belongs to the set of density operators on the input Hilbert space. Using this description for the hiding map, we see that the Kraus operators are given by $F_{kn}=\sqrt{p_{k}}\ketbra{k}{n}$ with $\sum_{kn}F_{kn}^\dagger F_{kn}=I$, where $\{\ket{n}\}$ and $\{\ket{k}\}$ are the basis states for the Hilbert space of the input system and the output system respectively. If the hiding map acts on a pure input state $\rho= \ketbra{\psi}{\psi}$ with $\ket{\psi} = \sum_m c_m \ket{m}$, then we have 
\begin{eqnarray}
\mathcal{E}(\rho)&=&\sum_{kn}F_{kn}\ketbra{\psi}{\psi}F_{kn}^\dagger\nonumber\\
&=&\sum_{kn}p_k\abs{\innerproduct{m}{\psi}}^2  \ketbra{k}{k}\nonumber\\
&=&\sum_{k}p_{k}\ketbra{k}{k}= \sigma, \label{equation:2}
\end{eqnarray}
where $\sigma$ is fixed for all $\rho$. Thus, the hiding map completely washes out the original information.

\subsection{Indefinite Causal Order}
One of the basic features of quantum mechanics is the superposition principle where a general state can be in a superposition of the basis states \cite{anton}. Extending this idea of superposition to the order of quantum channels applied on a state leads us to the concept of indefinite causal order or the notion of quantum switch \cite{comp}. It is a supermap that takes two quantum channels as inputs and outputs a new quantum channel which is a series concatenation of the two input channels in an indefinite order which is correlated with the state of a control qubit. Mathematically, if $\mathcal{N}_{1}$ and $\mathcal{N}_{2}$ are two quantum channels (acting on $\rho$) as inputs to the switch supermap $\mathcal{S}$ then the output channel $\mathcal{S}(\mathcal{N}_{1},\mathcal{N}_{2})$ acts as
\begin{equation}
    \mathcal{S}(\mathcal{N}_{1},\mathcal{N}_{2})(\omega)= \mathcal{N}_{2}\mathcal{N}_{1}(\innerproduct{0}{\omega|0}_{c}) + \mathcal{N}_{1}\mathcal{N}_{2}(\innerproduct{1}{\omega|1}_{c}), \nonumber
\end{equation}
where $\omega=(\rho\otimes\rho_{c})$, $\rho$ and $\rho_{c}$ are the states of the input system and the control qubit respectively, and $\innerproduct{i}{\omega|i}_{c}$ is the state of the input system conditioned on the outcome of a measurement in the basis $\{\ket{i}\}$ on the control qubit. If the control qubit is in the state $\ketbra{0}{0}_c$, $\mathcal{N}_{2}\mathcal{N}_{1}$ is applied on $\rho$ and if the control qubit is in the state $\ketbra{1}{1}$, $\mathcal{N}_{1}\mathcal{N}_{2}$ is applied on $\rho$.
If the control qubit is in a superposition of the states $\ket{0}_{c}$ and $\ket{1}_{c}$ then we see that a superposition of $\mathcal{N}_{2}\mathcal{N}_{1}$ and $\mathcal{N}_{1}\mathcal{N}_{2}$ is applied on $\rho$.\\\\
Let $\{K_{i}^{1}\}$ and $\{K_{i}^{2}\}$ be the Kraus operators of the channels $\mathcal{N}_{1}$ and $\mathcal{N}_{2}$ respectively. Then the Kraus operators for  $\mathcal{S}(\mathcal{N}_{1},\mathcal{N}_{2})$ are given by
\begin{equation}
 W_{ij}=K_{i}^{(2)}K_{j}^{(1)}\otimes\ketbra{0}{0}_{c} + K_{j}^{(1)}K_{i}^{(2)}\otimes\ketbra{1}{1}_{c} \nonumber \\
\end{equation}
and the channel acts on its input as
\begin{equation}
    \mathcal{S}(\mathcal{N}_{1},\mathcal{N}_{2})(\rho\otimes \rho_{c})= \sum_{i,j}W_{ij}(\rho\otimes \rho_{c})W_{ij}^\dagger,\label{equation:3}
\end{equation}
where $\sum_{ij} W_{ij}^{\dagger} W_{ij} = 1$.
The superposition of causal order has several applications in quantum information. It is well known that the Holevo information \cite{hol} of a constant channel such as the completely depolarising channel is zero. However, it was recently shown \cite{comm} that quantum switch allows a constant channel to have non-zero Holevo information where one can use two completely depolarising maps as quantum channels. It was found that the output is not completely depolarised but depends on the input state. In the subsequent section we consider the hiding map, of which the completely depolarising channel is a special case, in conjunction with the notion of indefinite causal order and show how to hide information in the bipartite correlation.

\section{Hiding map with Indefinite Causal Order}
  In this section we use indefinite causal order to apply two identical hiding maps $\mathcal{E}$ on a pure input state $\rho = \ketbra{\psi}{\psi}$. For the hiding map the Kraus operators of the switched channel are given by
 \begin{equation}
  W_{in,jm}=F_{in}^{(2)}F_{jm}^{(1)}\otimes\ketbra{0}{0}_{c} + F_{jm}^{(1)}F_{in}^{(2)}\otimes\ketbra{1}{1}_{c},   \label{equation:4}
 \end{equation}
  where $ F_{in}=\sqrt{p_{i}}\ketbra{i}{n}$.
 Let the control qubit be in a state $\rho_{c}= \ketbra{\phi}{\phi}_c$ where $\ket{\phi}_c= \sqrt{p} \ket{0} + \sqrt{1-p} \ket{1}$. Then using Eq.(\ref{equation:3}) we obtain

\begin{widetext}
\EQ{
\map S (\map E, \map E)\left(\rho \otimes \rho_c \right)&=&\sum_{ijnm} \Big(F_{in} F_{jm}\otimes \ketbra{0}{0}_c + F_{jm} F_{in}\otimes\ketbra{1}{1}_{{{c}}} \Big) \Big(\rho\otimes\rho_c\Big) \Big(F_{jm}^\dagger F_{in}^\dagger\otimes\ketbra{0}{0}_c+F_{in}^\dagger F_{jm}^\dagger\otimes\ketbra{1}{1}_{{{c}}} \Big) \nonumber \\
&=& \sum_{ijnm} \Big(F_{in} F_{jm} \rho F_{jm}^\dagger F_{in}^\dagger\otimes p\ketbra{0}{0}_{{{c}}}   +   F_{jm} F_{in} \rho F_{in}^\dagger F_{jm}^\dagger \otimes(1-p) \ketbra{1}{1}_{{{c}}} \nonumber \\
&& +   F_{in} F_{jm} \rho F_{in}^\dagger F_{jm}^\dagger\otimes \sqrt{p(1-p)} \ketbra{0}{1}_{{{c}}}+   F_{jm} F_{in} \rho F_{jm}^\dagger F_{in}^\dagger \otimes \sqrt{p(1-p)} \ketbra{1}{0}_{{{c}}}\Big) \nonumber \\
&=& \sum_{ijnm} \Big(F_{in}\sigma F_{in}^\dagger\otimes p\ketbra{0}{0}_{{{c}}}   +   F_{jm} \sigma F_{jm}^\dagger \otimes(1-p) \ketbra{1}{1}_{{{c}}} \nonumber \\
&& +   F_{in} F_{jm} \rho F_{in}^\dagger F_{jm}^\dagger\otimes \sqrt{p(1-p)} \ketbra{0}{1}_{{{c}}}+   F_{jm} F_{in} \rho F_{jm}^\dagger F_{in}^\dagger \otimes \sqrt{p(1-p)} \ketbra{1}{0}_{{{c}}}\Big) \nonumber \\
&=& \sum_{in} F_{in}\sigma F_{in}^\dagger\otimes \Big(p\ketbra{0}{0}_{{{c}}} + (1-p) \ketbra{1}{1}_{{{c}}}\Big)  + \sum_{ijnm}F_{in} F_{jm} \rho F_{in}^\dagger F_{jm}^\dagger\otimes\sqrt{p(1-p)} \Big(\ketbra{0}{1}_{{{c}}} + \ketbra{1}{0}_{{{c}}}\Big)\nonumber   \\
&=& \sigma\otimes \Big(p\ketbra{0}{0}_{{{c}}} + (1-p) \ketbra{1}{1}_{{{c}}}\Big)  + \sum_{ijnm}p_{i}p_{j}\ket{i}\delta_{nj}c_{m}c_{n}^{*}\delta_{im}\bra{j}\otimes\sqrt{p(1-p)} \Big(\ketbra{0}{1}_{{{c}}} + \ketbra{1}{0}_{{{c}}}\Big)\nonumber  \\
&=& \sigma\otimes \Big(p\ketbra{0}{0}_{{{c}}} + (1-p) \ketbra{1}{1}_{{{c}}}\Big)  + \sum_{ij}p_{i}p_{j}\ket{i}\innerproduct{i}{\psi}\innerproduct{\psi}{j}\bra{j}\otimes\sqrt{p(1-p)} \Big(\ketbra{0}{1}_{{{c}}} + \ketbra{1}{0}_{{{c}}}\Big)\nonumber  \\
&=& \sigma\otimes \Big(p\ketbra{0}{0}_{{{c}}} + (1-p) \ketbra{1}{1}_{{{c}}}\Big)  + \sigma\rho\sigma\otimes\sqrt{p(1-p)} \Big(\ketbra{0}{1}_{{{c}}} + \ketbra{1}{0}_{{{c}}}\Big), \label{equation:5}  }
\end{widetext}
where the first equality follows from Eq.(\ref{equation:4}). The second equality is the application of the Kraus operators on the joint system-control state. The third equality and the rest follow from the application of the hiding map on the input system and simplifying thereafter.

Thus, we can see that the joint system-control output state depends on the input state $\rho$ even though we have used the hiding map twice. If the control qubit is not in superposition, the output state of the joint system will not depend on the input. Also, the input information is neither in the system nor in the control and remains solely in the bipartite correlation between them. It may be noted that this correlation is clearly quantum since it is lost if the control decoheres in the computational basis. Thus, it is possible to hide information about the system in correlations if we add a control qubit which controls the order of the hiding maps acting on $\rho$. One may ask: Does this contradict the no-hiding theorem? The answer is no. If we consider the unitary version of the indefinite causal order, then we will still be able to find the original information in the ancilla subspace. The unitary version of the hiding map in the presence of indefinite causal order can be written as 
\begin{equation}
\ket{\psi} \otimes \ket{\psi}_c \otimes \ket{A} \rightarrow \sum_{ij} W_{ij} (\ket{\psi} \otimes \ket{\psi}_c) \otimes \ket{A_{ij}},\label{equation:6}
\end{equation}
where $\ket{A}$ is the initial state of the ancilla and $\ket{A_{ij} }$ are the mutual orthogonal basis in the ancilla Hilbert space. If we apply the no-hiding theorem, we can ascertain that the original information which is neither in the output state of the system nor in the 
output state of the control qubit, simply remains in the subspace of the ancilla Hilbert space (up to local unitary). Thus, the hiding map in the presence of indefinite causal order provides a new twist to hide information in the quantum correlation between two subsystems without violating the no-hiding theorem.\\

{\it Masking quantum information.--} The no-masking theorem states that an arbitrary pure quantum state cannot be masked unitarily. The process of masking quantum information is to keep information in the correlation, but not in the local subsystems. However, as we will show now, using two hiding maps in the presence of indefinite causal order we can map an arbitrary quantum state to a bipartite state such that the individual subsystems are independent of the initial state. The mapping is given by Eq.(\ref{equation:5}) where the state of the control qubit $\rho_{c}$ is fixed and is independent of the state of the input system $\rho_{s}$. In addition to the requirement that the global state contains information about the initial state $\rho_{s}$, the condition of masking says that $\Tr_{s|c}[\mathcal{S}(\mathcal{E},\mathcal{E})(\rho_{s}\otimes\rho_{c})]$ is independent of $\rho_{s}$. Now, taking the partial trace of the bipartite state from Eq.(\ref{equation:5}) we have the reduced states of the two subsystems as follows
\begin{widetext}
\EQ{
\Tr_{c}[\mathcal{S}(\mathcal{E},\mathcal{E})(\rho_{s}\otimes\rho_{c})] &=& \sigma, \label{equation:7}\\
\Tr_{s}[\mathcal{S}(\mathcal{E},\mathcal{E})(\rho_{s}\otimes\rho_{c})] &=& p\ketbra{0}{0}+(1-p)\ketbra{1}{1} + \Tr(\sigma\rho_{s}\sigma)\sqrt{p(1-p)}(\ketbra{0}{1}+\ketbra{1}{0}). \label{equation:8}
}.
\end{widetext}
We see that the reduced state of the first subsystem is independent of $\rho_{s}$ but the second one depends on it through  $\Tr(\sigma\rho_{s}\sigma) $. There are two ways to see how masking can be achieved perfectly. First, if we fix the output state as the maximally mixed state, i.e, $\sigma=I/d$ then $\Tr(\sigma\rho_{s}\sigma)=1/d^{2}$ and it is independent of the initial state $\rho_{s}$. 
Hence, the reduced states of both the subsystems do not have any information about the input state $\rho_{s}$ and the conditions for masking are satisfied. Therefore, an arbitrary quantum state can be masked by this map. Second, let us put no restriction on $\sigma$, then the set of quantum states $\{\rho_{1},\rho_{2},...\}$ can be masked if $\Tr(\sigma\rho_{i}\sigma)= K $, where $K$ is independent of $i$. For any two states from the set, our condition then demands that $\Tr(\sigma\rho_{i}\sigma)=\Tr(\sigma\rho_{j}\sigma)$. For simplicity. let us consider the case of pure states and take $\rho_{i}=\ketbra{\psi}{\psi}$ and $\rho_{j}=\ketbra{\phi}{\phi}$. It implies then that $\innerproduct{\psi}{\sigma^{2}|\psi}=\innerproduct{\phi}{\sigma^{2}|\phi}$. Since there always exists a unitary $U$ such that $\ket{\phi}=U\ket{\psi}$, we have $\innerproduct{\psi}{\sigma^{2}|\psi}=\innerproduct{\psi}{U^{\dagger}\sigma^{2}U|\psi}$ which is true for arbitrary $\ket{\psi}$. Therefore, we get
\EQ{
U^{\dagger}\sigma^{2}U&=&\sigma^{2},\nonumber\\
\comm{U}{\sigma^{2}}&=&0.\label{equation:9}
}
In the eigenbasis $\{\ket{k}\}$ of $\sigma^{2}$ then $U$ can be written as $U=\sum_{k}a_{k}\ketbra{k}{k}$ with  $a_{k}=\exp{-i\phi_{k}}$. Since $\ket{\phi}=U\ket{\psi}$, now expanding the two state vectors in the basis $\{\ket{k}\}$ we have
\EQ{
\sum_{k}b_{k}\ket{k}&=&\sum_{km}\exp{-i\phi_{k}}c_{m}\ket{k}\innerproduct{k}{m}\nonumber\\
b_{k}&=&\exp{-i\phi_{k}}c_{k}  \hspace{0.4cm}\forall\hspace{0.1cm} k.\label{equation:10}
}
Since the equation above is true for any two states of the set, therefore, the set of all those states whose expansion coefficients in the $\{\ket{k}\}$ basis differ by at most a phase factor can be masked by this map. Thus, we see that masking of quantum information is allowed through indefinite causal order, i.e., information about the input state can remain in the correlation of the bipartite output state, although the mapping is not unitary but general CPTP map. \\

{\it Preserving information.--} Since the original information about the input now remains in the correlation, one can ask what is the fidelity between the original state and the conditional state after we perform a measurement on the control qubit. Now, measuring the control qubit in $\ket\pm$ basis and tracing it out we get the conditional output states for the system as
\begin{equation}
\bra\pm\mathcal{S}(\mathcal{E,E})(\rho \otimes \rho_{c})\ket\pm= \frac{\sigma}{2} \pm \sqrt{p(1-p)}\sigma\rho\sigma = \sigma_{\pm}. \label{equation:11}
\end{equation}
The normalised conditional states are given by $\rho_{\pm} = \sigma_{\pm} / \Tr {\sigma_{\pm}} $.
To check the amount of quantum information that is preserved under the hiding map with the assistance of quantum switch we can use the concept of fidelity \cite{rich}. The fidelity between two states $\ket{\psi}$ and $\rho$ is defined as $F=\innerproduct{\psi}{\rho|\psi}$ where $\rho$ is normalised. Using this definition for our case we find
\begin{equation}
F(\rho_{+})= \innerproduct{\psi}{\rho_{+}|\psi}
= \Big(\frac{\frac{1}{2}+\sqrt{p(1-p)}\innerproduct{\psi}{\sigma|\psi}}{\frac{1}{2}+\sqrt{p(1-p)} \innerproduct{\psi}{\sigma^2|\psi}}\Big)\innerproduct{\psi}{\sigma|\psi}. \label{equation:12}
\end{equation}

Now, we can show that the fidelity of the conditional state $\rho_+$ with respect to the original input $\rho$ is greater than or equal to the fidelity of the bleached out state $\sigma$ with respect to the original input $\rho$. Note that $F(\sigma)= \innerproduct{\psi}{\sigma|\psi} =\sum_{k}p_{k}\abs{\innerproduct{\psi}{k}}^2$ and $\innerproduct{\psi}{\sigma^2|\psi} =\sum_{k}p_{k}^2\abs{\innerproduct{\psi}{k}}^2$. Since $0\leq p_{k}\leq1 $ $\forall$ $k$, we have  $p_{k}^2\leq p_{k}$ and hence  $\innerproduct{\psi}{\sigma^2|\psi}\leq\innerproduct{\psi}{\sigma|\psi}$. Therefore, we see from Eq.(\ref{equation:12}) that $F(\alpha_{+})\geq F(\sigma)$.  Since $\sigma$ is a fixed state independent of $\psi$, $F(\sigma)$ should not take a value more than the classically attainable fidelity, otherwise one may argue that there is some original quantum information in the bleached out state $\sigma$. For a 
single qubit, if the fidelity between the output and the input state is more than $\frac{2}{3}$ (highest classically attainable value), then we can say that the output state has some 
quantum information about the input state \cite{massar}. Now, assume that $F(\sigma) = \frac{2}{3}$, then our result implies that the fidelity 
$F(\rho_+)$ is more than $\frac{2}{3}$, thus capturing some quantum information about the input state and thus preserving the quantum information.\\
 
{\it Preserving quantum coherence.--} How to preserve quantum coherence \cite{alex,loyd} under noisy channel is an important question in the study of 
decoherence process \cite{max,max2,zurek}. If an arbitrary state is sent through the hiding map, the coherence of the initial state is completely lost.
Note that in the basis $\{\ket{k}\}$, the bleached out state $\sigma=\sum_{k}p_{k}\ketbra{k}{k}$ is an incoherent state since it is diagonal in that basis. For an arbitrary state $\rho$, the $l_{1}$ norm of coherence \cite{baum} is given by the sum of the absolute value of the off-diagonal elements of $\rho$ i.e., $C_{l_{1}}(\rho)=\sum_{i\neq j}\abs{\rho_{ij}}$ where $\rho_{ij}$ denotes $ij^{th}$ element of $\rho$. Using this definition we see that the initial coherence of the input state $\ket{\psi}(=\sum_{i}c_{i}\ket{i})$ is $C_{l_{1}}(\psi) = \sum_{i\not=j} |c_i||c_j|$ and the coherence of the final state under hiding map is given as $C_{l_{1}}(\sigma)=0$. Here, we will show that one can use indefinite causal order to preserve some non-zero coherence. We find that the coherence of the normalised conditional states  $\rho_{\pm}=\sigma_{\pm}/Tr(\sigma_{\pm})$ is given as
\EQ{
C_{l_{1}}(\rho_{\pm})=\frac{\sum_{k\neq l}\sqrt{p(1-p)}p_{k}p_{l}\abs{c_{k}}\abs{c_{l}}}{1/2\pm\sqrt{p(1-p)}\sum_{m}\abs{c_{m}}^2p_{m}^{2}},\label{equation:13}
}
where $c_{k}=\innerproduct{k}{\psi}$. Thus, the single use of the hiding map destroys coherence whereas superposition of two hiding maps with indefinite causal order helps to preserve quantum coherence of the initial state.\\

{\it Work extraction.--} In thermodynamics information about a physical system can be used to extract work from it \cite{ben,land,goold}. Given a quantum state $\rho$ we can define extractable work from it as a function of the entropy of the system as \cite{jon}
\EQ{
W({\rho})=\log d -S({\rho}), \label{equation:14}
}
where $d$ is the dimension of the system. For simplicity we discuss here the case of $d=2$.
Consider the output of the hiding map, $\sigma$, its entropy is $S(\sigma)=-\sum_{k}p_{k}\log p_{k}$ where $p_{k}$ are the eigenvalues of $\sigma$. If we take $p_{k}=1/2$ $\forall k$ then $S(\sigma)=\log 2$ which implies $W(\sigma)=0$, thus no work can be extracted from it. Now consider the conditional states $\rho_{\pm}$, the entropy can be written as $S(\rho_{\pm})=-\sum_{k}\lambda_{k_{\pm}}\log\lambda_{k_{\pm}}$ where $\lambda_{k_{\pm}}$ are the eigenvalues of $\rho_{\pm}$. Using Eq.(\ref{equation:11}) we see that the eigenvalues are
\EQ{
\lambda_{1_{\pm}}=\frac{1}{2}(1+x_{\pm}),\nonumber\\
\lambda_{2_{\pm}}=\frac{1}{2}(1-x_{\pm}),\nonumber\\\label{equation:15}
}
where 
\EQ{
x_{\pm}=\sqrt{1-\frac{4p_{0}p_{1}\Big(1\pm2\sqrt{p(1-p)}\sum_{m}\abs{c_{m}}^2p_{m}\Big)}{\Big(1\pm2\sqrt{p(1-p)}\sum_{m}\abs{c_{m}}^2p_{m}^2\Big)^2}}.\nonumber
}

We can now show that indefinite causal order of two hiding maps can give non-zero extractable work $W$ even when a single hiding map gives $W(\sigma)=0$. Since the two eigenvalues are different from $1/2$ it implies that $\rho_{\pm}$ is not completely random which in turn implies that $S(\rho_{\pm})<\log 2$. Therefore, using Eq.(\ref{equation:14}), we see that $W(\rho_{\pm})>0$. For example, if we take $c_{k}=1/\sqrt{2}, p_{k}=1/2$  $\forall k$ and $p=1/2$,
then using Eq.(\ref{equation:14}) it can be checked that $W(\rho_{+})=0.020$ and $W(\rho_{-})=0.057$. This clearly brings out the usefulness of indefinite causal order in work extraction.

\section{Hiding map, Indefinite Causal Order and entangled input}
In the previous section, we have shown that with the help of indefinite causal order one can preserve quantum information and quantum coherence. Here, we ask can we also preserve entanglement under superposition of hiding maps. Let us consider a pure bipartite entangled state $\rho=\ketbra{\Psi}{\Psi}$, where $\ket{\Psi}= \sum_{i}c_{i}\ket{ii}$. 
If we apply hiding map to one of the subsystems (say, second) the entanglement between the two subsystems is lost and the joint output state is given as
\EQ{
 \mathcal{E}(\rho)= \sum_{kn}F_{kn}\ketbra{\Psi}{\Psi}F_{kn}^\dagger\nonumber\\
 =\sum_{i}\abs{c_{i}}^{2}\ketbra{i}{i}\otimes\sigma.\label{equation:16}
}
We see that the output is the tensor product of decohered state of the input system and a fixed state independent of the input system. The no-hiding theorem says that this lost information has gone in the remainder of the ancilla. The entanglement between the two subsystems will now be transferred as entanglement between the first subsystem and another system which may be part of the ancilla.
Now, if we apply the hiding map in the presence of indefinite causal order, then can we protect this information? That is can we preserve some amount of entanglement by applying the hiding maps in superposition? We will see that the answer to this question is no. Even though in the larger Hilbert space,  the unitary map that will implement the quantum switch can preserve entanglement (by transferring to another subsystem), at the level of input-control and output-control system, we cannot preserve entanglement between the original two subsystems. The conditional output state does not have any entanglement across the bipartition.

To see this we apply the superposition of two hiding maps via quantum switch on the second subsystem of $\ket{\Psi}$. Taking the same $\rho_{c}$ as in the previous subsection and using Eq.(\ref{equation:3}) we obtain

\begin{widetext}
\EQ{
\map S (\map E, \map E)\left(\rho \otimes \rho_c \right)&=& \sum_{ijnm} \Big(F_{in} F_{jm} \rho F_{jm}^\dagger F_{in}^\dagger\otimes p\ketbra{0}{0}_{{{c}}}  +  F_{jm} F_{in} \rho F_{in}^\dagger F_{jm}^\dagger\otimes (1-p) \ketbra{1}{1}_{{{c}}}  \nonumber \\
&& + F_{in} F_{jm} \rho F_{in}^\dagger F_{jm}^\dagger\otimes \sqrt{p(1-p)} \ketbra{0}{1}_{{{c}}} + F_{jm} F_{in} \rho F_{jm}^\dagger F_{in}^\dagger\otimes \sqrt{p(1-p)} \ketbra{1}{0}_{{{c}}} \Big) \nonumber \\
&=&\sum_{i}\abs{c_{i}}^{2}\ketbra{i}{i}\otimes\sigma\otimes \Big(p\ketbra{0}{0}_{{{c}}}+(1-p) \ketbra{1}{1}_{{{c}}}\Big) + (I\otimes\sigma)\rho(I\otimes\sigma)\otimes \sqrt{p(1-p)} \Big(\ketbra{0}{1}_{{{c}}} + \ketbra{1}{0}_{{{c}}}\Big), \nonumber}
\end{widetext}
where we used $\sum_{jm}F_{jm}\rho F_{jm}^\dagger= \sum_{i}\abs{c_{i}}^{2}\ketbra{i}{i}\otimes\sigma$ and $\sum_{injm}F_{jm}F_{in}\rho F_{jm}^{+}F_{in}^{+}= (I\otimes\sigma)\rho(I\otimes\sigma)$.
To use the advantage of quantum switch, again measuring the control qubit in $\ket\pm$ basis and tracing it out we get the conditional output states as follows
\begin{widetext}
\begin{equation}
\bra\pm S(\mathcal{E,E})(\rho \otimes \rho_{c})\ket\pm= \sum_{i}\abs{c_{i}}^{2}\ketbra{i}{i}\otimes\frac{\sigma}{2} \hspace{.2cm}\pm\hspace{.2cm}\sqrt{p(1-p)}(I\otimes\sigma)\rho(I\otimes\sigma) = \alpha_{\pm}.\label{equation:17}
\end{equation}
\end{widetext}

 We can now show that the hiding map with the assistance of indefinite causal order does not preserve any entanglement. For simplicity, we consider the case of two qubit entangled state $\ket{\Psi} = c_0 \ket{00} + c_1 \ket{11}$, as the input. If we apply the hiding map on the second qubit, the reduced state will transform to the state  $\sigma=p_{0}\ketbra{0}{0}+p_{1}\ketbra{1}{1}$, which is a fixed state. 
 Using the form of $\ket{\Psi}$ and $\sigma$ in Eq.(\ref{equation:17}) the conditional unnormalized output states $\alpha_{\pm}$ can be written as
\begin{widetext}
\EQ{
\alpha_{\pm}&=&\abs{c_{0}}^{2}\frac{p_{0}}{2}\Big(1\pm 2\sqrt{p(1-p)}p_{0}\Big)\ketbra{00}{00} \pm\sqrt{p(1-p)}p_{1}p_{0}c_{0}c_{1}^{*}\ketbra{00}{11}+ \abs{c_{0}}^{2}\frac{p_{1}}{2}\ketbra{01}{01} + \abs{c_{1}}^{2}\frac{p_{0}}{2}\ketbra{10}{10} \nonumber \\
&& \pm\sqrt{p(1-p)}p_{1}p_{0}c_{1}c_{0}^{*}\ketbra{11}{00}+\abs{c_{1}}^{2}\frac{p_{1}}{2}\Big(1\pm 2\sqrt{p(1-p)}p_{1}\Big)\ketbra{11}{11}.\label{equation:18}
}\\
\end{widetext}

We can now use the PPT (Positive Partial Transpose) criteria \cite{asher,horo96} to see that there remains no entanglement in $\alpha_{\pm}$. The criteria states that if a bipartite state $\rho_{AB}$ is separable then all of the eigenvalues of the partially transposed state $\rho_{AB}^{T_{A}}$ i.e., $(T_{A} \otimes I_{B})\rho_{AB}$ are non-negative, where $T_{A}$ is the transpose operator acting on the first subsystem and $I_{B}$ is the identity operator on the second subsystem.

Since PPT criteria is both necessary and sufficient for the case of 2 qubits, therefore $\alpha_{\pm}$ will be separable if all of the eigenvalues of $(T_{A} \otimes I_{B})\alpha_{\pm}$ are non-negative. The four eigenvalues of $(T_{A} \otimes I_{B})\alpha_{\pm}$ can be written as
\begin{eqnarray}
\lambda_{1{\pm}}&=&\abs{c_{0}}^{2}\frac{p_{0}}{2}\Big(1\pm 2\sqrt{p(1-p)}p_{0}\Big),\nonumber\\
\lambda_{2{\pm}}&=&\abs{c_{1}}^{2}\frac{p_{1}}{2}\Big(1\pm 2\sqrt{p(1-p)}p_{1}\Big),\nonumber\\
\lambda_{3{\pm}}&=&\frac{1}{2} \Big[x+\sqrt{x^2-\abs{c_{0}}^2\abs{c_{1}}^2p_{0}p_{1}\Big(1-4p(1-p)p_{0}p_{1}\Big)} \Big], \nonumber\\
\lambda_{4{\pm}}&=&\frac{1}{2} \Big[x-\sqrt{x^2-\abs{c_{0}}^2\abs{c_{1}}^2p_{0}p_{1}\Big(1-4p(1-p)p_{0}p_{1}\Big)} \Big],\nonumber\\\label{equation:19}
\end{eqnarray}
where $x= \frac{1}{2}\Big(\abs{c_{0}}^2p_{1} +\abs{c_{1}}^2p_{0}\Big)$.
Using normalization conditions on the probabilities and the coefficients, it can be checked that all of the four eigenvalues of the transformed $\alpha_{\pm}$ are non-negative. Thus, using the PPT criteria this implies that $\alpha_{\pm}$ is a separable state. Therefore, two uses of the hiding map 
in the presence of indefinite causal order cannot preserve entanglement. Now, if Alice and Bob share an entangled pair and during the entanglement distribution Bob's qubit passes through the hiding map in the presence of indefinite causal order, then the coherent information 
can never be positive. This shows that the indefinite causal order cannot help in sending distinct quantum information across two uses of the noisy channel.\\

{\it Entanglement Fidelity.--} Here, we show that even though entanglement between the pair $AB$ is destroyed, 
the entanglement fidelity \cite{schumacher} is non-zero under superposition of two hiding maps.
This shows that the entanglement fidelity  is not a good measure of entanglement preservation of a state, with a part of the system undergoing some general CPTP map ${\cal E}$ and the other part being dynamically isolated. If we start with a pure entangled state $\ket{\Psi}_{AB}$ of a pair of systems $AB$, and send the subsystem $B$ through a general quantum channel $\mathcal{E}$, then the entanglement fidelity can be defined as 
$$F_{e}= _{AB}\bra{\Psi}\rho_{AB'}\ket{\Psi}_{AB}, $$
where $\rho_{AB'} = I_A\otimes {\cal E}_B (\ket{\Psi}_{AB} \bra{\Psi}) $. Physically, it is the probability that the final state $\rho_{AB'}$ would agree with the initial state $\ket{\Psi}_{AB}$. Even though it was initially thought of as a measure of how well we can preserve entanglement between $A$ and $B$ when the subsystem $B$ undergoes a noisy evolution, it does not capture this aspect. It has more to do with a measure of how much the system $B$ undergoes disturbance under a dynamical map ${\cal E}$\cite{schumacher}.

\begin{figure}[ht]
    \centering
    \includegraphics[width=8.5cm]{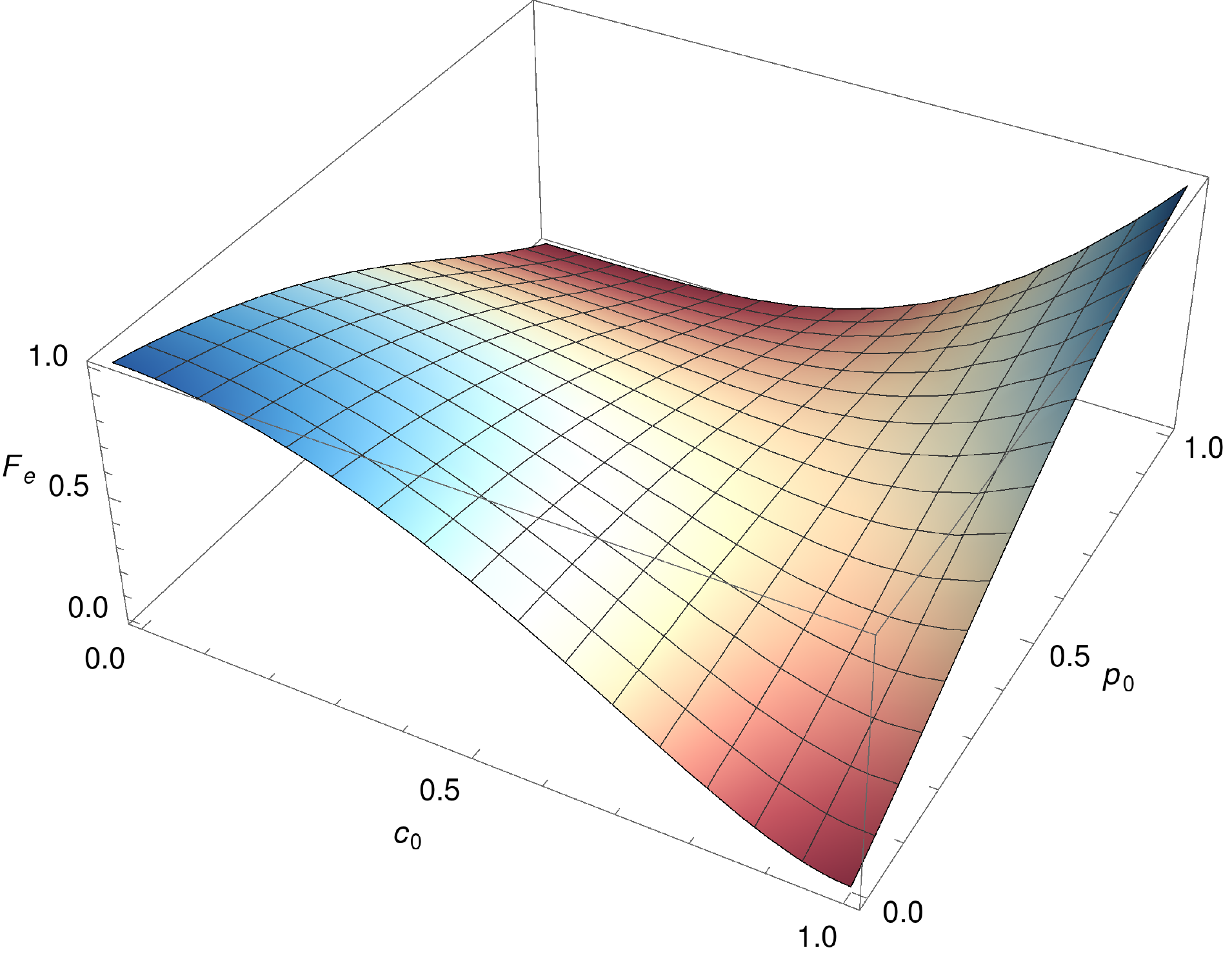}
    \caption{Entanglement fidelity $F_e$, plotted as a function of $c_0$ and $p_0$, when the final state of the system $AB'$ is $\rho_{+}$ and the control qubit $\ket{\phi}_c$ is initialised as the maximally coherent state $\ket{+}$.}
    \label{fig:ef}
\end{figure}

In the case considered above, let the initial entangled state be $\ket{\Psi }_{AB} = c_0 \ket{00} + c_1 \ket{11}$ and the transformed state with normalization, under the application of two hiding maps with indefinite causal order on $B$, can be written using Eq.(\ref{equation:18}) as $\rho_{\pm}=\alpha_{\pm}/\Tr{\alpha_{\pm}}$. For simplicity, we take the initial state of the control qubit as $\ket{\phi}_c=\ket{+}$, i.e., $p=0.5$. Let us consider the case when the measurement on the control qubit gives $\ket{+}$ as outcome, the corresponding state of $AB'$ is then $\rho_{+}$. The entanglement fidelity of this process can then be calculated to give
\begin{widetext}
\EQ{
F_e&=& {}_{AB}\bra{\Psi }\rho_{+}\ket{\Psi}_{AB}\nonumber\\
&=&\frac{2-4\abs{c_0}^{2}(1-p_0)^{2}-(3-p_0)p_{0}+2\abs{c_0}^{4}(1-2p_{0}(1-p_0))}{2-(2-p_0)p_{0}-\abs{c_0}^{2}(1-2p_0)}.
\nonumber\\\label{equation:20}
}
\end{widetext}

We now plot the entanglement fidelity $F_e$ as a function of $c_0$ and $p_0$ where we take $c_0$ to be real for simplicity (this condition does not restrict our result in any way). We see from Fig.~\ref{fig:ef} that $F_e$ approaches unity for arbitrary values of $c_0$ and $p_0$ but never reaches it unless $(c_0,p_0)=(0,0)$ or $(1,1)$. In both of these cases the initial state $\ket{\Psi}_{AB}$ is separable and the final state $\rho_{+}$ overlaps it completely as can be seen by Eq.(\ref{equation:18}). Similarly $F_e$ vanishes for $(c_0,p_0)=(0,1)$ or $(1,0)$ because in both of these cases $\ket{\Psi}_{AB} $ and $\rho_{+}$ are orthogonal to each other as is evident by Eq.(\ref{equation:18}).

However, the problem arises when we analyse the cases where $F_e$ approaches $1$ but never reaches it. For such cases the value of $F_e$ then implies that the final state is very close to the initial state. In other words, for some non-zero values of $c_0$ and $p_0$ the entanglement  present in the initial state $\ket{\Psi}_{AB}$ is preserved in the final state $\rho_{+}$ to a great extent. This is in stark contrast to what we observed previously in this section using PPT criteria for $\rho_{+}$, where we found that $\rho_{+}$ is a  separable state for all values of $c_0$ and $p_0$. Therefore, we should not identify the entanglement fidelity as a measure of entanglement preservation when one part of the system remains dynamically isolated and the other goes through a quantum process. This kind of behavior of the entanglement fidelity has also been reported before \cite{xiang}. In addition, originally it was noted by Schumacher \cite{schumacher} that the entanglement fidelity $F_e$ can be lowered by a local unitary on $B$ as well as by information exchange with the environment where as entanglement in the system $AB$ is independent of the internal dynamics of $B$. 
Using this observation one can define the modified entanglement fidelity as
\EQ
{
F'_{e}=\max_{U_{B}}{}_{AB}\bra{\Psi}(I_{A}\otimes U_{B})\rho_{AB'}(I_{A}\otimes U_{B})^{\dagger}\ket{\Psi}_{AB}\nonumber\\\label{equation:21}
}
which works very well for some cases where the usual entanglement fidelity fails \cite{xiang}.
However, since $F'_e\geq F_e$ it is clear that even the modified form of entanglement fidelity $F'_e$ has the same problems as $F_e$ in the example considered above when we encounter the case of it approaching $1$ but not reaching it.\\

\vskip .5cm

\section{Conclusions}
In this paper, we have discussed the bleaching process for quantum information and how quantum information is apparently lost under the bleaching process. However, the no-hiding theorem proves that the lost information can always be recovered, in principle, from the ancilla Hilbert space. In addition to the no-hiding theorem, the no-masking theorem in quantum mechanics is another no-go result which rules out the possibility of storing information in the bipartite correlation by a 
unitary encoding mechanism. However, when we allow superposition of two hiding maps in the presence of indefinite causal order, then it is possible to keep the information about the input state
in the bipartite correlation under some condition. Though, this does not contradict the
no-hiding and the no-masking theorems. If we look at the unitary version of the hiding map with indefinite causal order, then information can be recovered from the ancilla Hilbert space in accordance with the no-hiding theorem. Nevertheless, indefinite causal order provides a mechanism to store arbitrary state (both pure and mixed) in the bipartite
correlation using non-unitary encoding mechanism. We have shown that for a single input state, the superposition of two hiding maps in the presence of indefinite causal order can preserve some quantum information, helps in work extraction and preserves quantum coherence. However, if we have an entangled state as input and apply the superposition of two hiding maps in the presence of indefinite causal order on one half of the entangled pair, then the entanglement is completely lost. Thus, even though quantum information and quantum coherence are partially immune to noise in the presence of indefinite causal order, quantum entanglement is not. We have also investigated the behavior of the entanglement fidelity under superposition of two hiding maps and found that the former does not vanish even though there does not remain any entanglement.

The possibility of storing information only in the correlations can have several important applications in quantum information as well as in other areas of physics. This can have some implications even for the 
black hole information loss. For example, if we think of black hole as a thermalising 
quantum channel,
then may be if we superpose two black holes with indefinite causal order,
then the information about the input state may be completely hidden in the correlation. This
may provide an alternate resolution to information loss. It may be
possible that when space time geometry is too much distorted, two black holes may be in an indefinite
causal orders which may be naturally happening. Then, information is neither inside nor outside the black hole but completely in the correlation. We believe that our result can have interesting applications quantum computing, quantum qubit commitment, quantum communication, secret sharing, black hole information loss and in other physical processes as well.

\bibliography{ref}

\end{document}